\documentclass[sigconf]{acmart}
\usepackage{amsmath}    
\usepackage{amsfonts}

\usepackage{soul}
\usepackage{graphicx}

\usepackage{caption, subcaption}
\usepackage{natbib}
\usepackage{hyperref}
\hypersetup{
	colorlinks=true,
}
\usepackage{multirow}
\usepackage{algorithm}
\usepackage{algorithmic}
\usepackage{makecell}
\usepackage{url}

\usepackage{footnote}

\usepackage{textcomp}
\usepackage{xcolor}

\usepackage{booktabs}
\usepackage{colortbl}

\usepackage{mdframed}

\usepackage{tablefootnote}
\usepackage{pifont}
\definecolor{myblue}{HTML}{E0F3F7}
\definecolor{darkgreen}{RGB}{0, 100, 0} 
\definecolor{mygray}{RGB}{214, 214, 214} 

\AtBeginDocument{%
  }

\setcopyright{acmlicensed}
\copyrightyear{2026}
\acmYear{2026}
\acmDOI{XXXXXXX.XXXXXXX}
\acmConference[26]{Conference}{Jan 1--10,
  2026}{XX, XX}
\acmISBN{978-1-4503-XXXX-X/2018/06}

\usepackage{microtype}
\begin{document}


\title{TICoder: A Repository-Level Code Generation Framework with\\Test-Driven Planning and Implementation-Aware Reuse}

\author{Siyu Nan}
\authornote{Both authors contributed equally to this research.}
\email{siyunan@whu.edu.cn}
\affiliation{%
  \institution{Wuhan University}
  \city{Wuhan}
  \state{Hubei}
  \country{China}
}

\author{Yaling Luo}
\authornotemark[1]
\email{luoyaling@whu.edu.cn}
\affiliation{%
  \institution{Wuhan University}
  \city{Wuhan}
  \state{Hubei}
  \country{China}
}

\author{Jian Wang}
\authornote{Corresponding authors.}
\email{jianwang@whu.edu.cn}
\affiliation{%
  \institution{Wuhan University}
  \city{Wuhan}
  \state{Hubei}
  \country{China}
}

\author{Neng Zhang}
\email{nengzhang@ccnu.edu.cn}
\affiliation{%
  \institution{Central China Normal University}
  \city{Wuhan}
  \state{Hubei}
  \country{China}
}

\author{Bing Li}
\authornotemark[2]
\email{bingli@whu.edu.cn}
\affiliation{%
  \institution{Wuhan University}
  \city{Wuhan}
  \state{Hubei}
  \country{China}
}

\renewcommand{\shortauthors}{Trovato et al.}

\begin{abstract}
  Repository-level code generation with Large Language Models (LLMs) remains challenging, primarily due to complex dependencies and limited context windows. Recent approaches adopt retrieval-augmented generation (RAG) and the planning mechanism to reuse potential callee functions in the repository.
  However, these approaches often suffer from two limitations: lack of test-driven behavioral guidance during planning and overlooking the implementation logic embedded in repository code during reuse. As a result, generated plans may not align with expected behaviors, and retrieved functions may not be effectively reused.
  In this paper, we propose TICoder, a novel repository-level code generation framework that improves both planning and reuse. TICoder introduces a test-driven iterative planning mechanism that leverages test cases as behavioral specifications to refine implementation steps. Furthermore, TICoder employs an implementation-aware code reuse strategy, which retrieves potential callee functions using a dual-view similarity that captures both functional and implementation aspects. We then identify relevant usage patterns through a dual-stage selection strategy, combining structure-based clustering and perplexity-based filtering. 
  We conduct extensive experiments on widely used repository-level code generation benchmarks with various LLMs. Experimental results demonstrate that TICoder outperforms state-of-the-art (SOTA) methods, achieving an average improvement of 11.52\%.

\end{abstract}



\received{20 February 2007}
\received[revised]{12 March 2009}
\received[accepted]{5 June 2009}

\maketitle

\section{Introduction}\label{Introduction}
Code generation is a critical task in modern software development, aiming to bridge natural language (NL) requirements with executable code \citep{CodeGenSurvey2026ASE, CodeGenSurvey2024ARXIV, CodeGenSurvey2025ESE, CodeGenSurvey2024TOSEM}. Recently, LLMs have demonstrated remarkable capabilities in generating standalone code snippets from NL requirements. However, they still struggle with repository-level code generation, a more realistic development scenario that requires understanding and leveraging the code repository. This task is particularly challenging due to complex dependencies and the limited context window of LLMs \citep{REPORACGSurvey2025ARXIV, CATCODER, HSREPOCodeUnderstand, REPOCodeGenTool}.

To mitigate these challenges, existing approaches adopt the RAG framework \citep{REPORACGSurvey2025ARXIV, CATCODER, RepoCoder, RLCoder}, where similar code snippets are retrieved from the repository and provided as contextual information to guide LLMs during generation. However, retrieving similar code snippets does not always improve generation performance and may even negatively affect it, because there is no guarantee that a functionally similar implementation corresponding to the NL requirement exists within the repository \citep{AllianceCoder}. In real-world development scenarios, developers often implement new functionality by composing multiple partially related functions rather than directly reusing a single similar code snippet.

To address this limitation, several recent studies propose planning-based repository-level Retrieval-Augmented Code Generation (RACG) approaches \citep{AllianceCoder, CodePlan, self-planning}. These approaches first generate a sequence of implementation steps for the task based on the NL requirement, and then retrieve potential callee functions from the repository. By decomposing the NL requirement into substeps, these approaches improve the discoverability of reusable repository functions and have shown promising performance in repository-level code generation. Despite these advances, we observe that existing approaches still suffer from two key limitations.

\noindent \textbf{L1. Lack of test-driven behavioral guidance in planning.}

Existing approaches \citep{AllianceCoder, CodePlan, self-planning} typically derive implementation steps directly from NL requirements. However, NL requirements in real-world development are often vague or incomplete, making it difficult to produce reliable implementation steps. In practice, developers frequently rely on test cases to clarify expected behaviors and guide implementation. While recent test-driven code generation methods \citep{TDDLLMCG, TDDBenchLLMCG, TDDCICG} leverage test cases to improve the performance of generated code, they mainly apply them in the generation stage, leaving their potential for guiding the planning process unexplored.

\noindent \textbf{L2. Lack of code-level implementation awareness in reuse.}

Existing planning-based RACG approaches \citep{AllianceCoder, CodePlan, self-planning} mainly capture the implementation steps at the requirement level while overlooking the implementation logic embedded in the code. During retrieval, potential callee functions are typically identified based on semantic similarity between implementation steps and function descriptions, while the execution logic of the code is ignored. Additionally, during generation, retrieved code snippets are often incorporated directly into prompts without understanding how these functions are actually used in the repository. Consequently, precise retrieval and effective reuse of potential callee functions in the repository remain challenging.

To address the aforementioned limitations, we propose TICoder, a novel repository-level code generation framework that improves both planning and reuse.
First, we introduce a test-driven iterative planning mechanism that incorporates test cases to guide the generation of implementation steps (addressing \textbf{L1}). Instead of deriving plans solely from NL requirements, TICoder leverages test cases as behavioral specifications to refine planning results. Specifically, the framework employs a judge-and-reflection planning process in which an \emph{LLM-as-a-planner} generates implementation steps, and an \emph{LLM-as-a-judge} evaluates them, enabling iterative refinement of planning results.
Second, we introduce an implementation-aware grounding strategy consisting of dual-view callee function retrieval and dual-stage usage pattern selection (addressing \textbf{L2}). The retrieval module identifies potential reusable functions by jointly considering functional similarity and implementation similarity, enabling more precise discovery of potential callee functions. The selection module further extracts relevant usage patterns through structure-based clustering and perplexity-based filtering.
Finally, TICoder performs enhanced code generation by integrating NL requirements, test cases, retrieved callee functions, and selected usage patterns as contextual information for LLMs, enabling LLMs to generate more effective code in repository-level code generation tasks. Table \ref{table_approaches} presents the comparison between TICoder and several representative repository-level code generation approaches.

We evaluate TICoder on two widely-used repository-level code generation datasets, CoderEval \citep{CoderEval} and DevEval \citep{DevEval}, utilizing three LLMs, GPT-4o-mini \citep{GPT}, DeepSeek-V3 \citep{DeepSeek}, and Qwen2.5-Coder \citep{Qwen2.5-Coder}, as backbones. The results show that TICoder outperforms the best-performing baseline across all backbones, achieving up to 11.52\% improvement on average.

In summary, the contributions are as follows:
\begin{itemize}
  \item We propose TICoder, a novel repository-level code generation framework with test-driven iterative planning and implementation-aware code reuse.

  \item We introduce an implementation-aware repository code reuse strategy through callee function retrieval based on dual-view similarity and usage pattern selection based on a dual-stage strategy.

  \item We conduct extensive experiments to evaluate TICoder. The results demonstrate that TICoder outperforms SOTA baselines and adapts to various LLMs. 
\end{itemize}

\begin{table}[t]
	\small
	\caption{Comparison between TICoder and existing repository-level code generation methods, including whether the method adopts a planning mechanism, utilizes test cases, improves retrieval, and exploits usage patterns.}
	\label{table_approaches}
	\vspace{-7pt}
	\centering
	\renewcommand{\arraystretch}{0.8}
	\setlength{\tabcolsep}{6pt}
	\begin{tabular}{ccccc}
		\toprule
		\textbf{Approaches}&\textbf{Planning}&\makecell{\textbf{Test}\\\textbf{Cases}}&\textbf{Retrieval}&\makecell{\textbf{Usage}\\\textbf{Patterns}}\\\midrule
		A$^3$Codgen \citep{A3CodGen}&\textcolor{red}{\ding{55}}&\textcolor{red}{\ding{55}}&\textcolor{darkgreen}{\ding{51}}&\textcolor{red}{\ding{55}}\\
		
		AllianceCoder \citep{AllianceCoder}&\textcolor{darkgreen}{\ding{51}}&\textcolor{red}{\ding{55}}&\textcolor{red}{\ding{55}}&\textcolor{red}{\ding{55}}\\
		
		CodePlan \citep{CodePlan}&\textcolor{darkgreen}{\ding{51}}&\textcolor{red}{\ding{55}}&\textcolor{red}{\ding{55}}&\textcolor{red}{\ding{55}}\\
		
		RepoScope \citep{RepoScope}&\textcolor{red}{\ding{55}}&\textcolor{red}{\ding{55}}&\textcolor{red}{\ding{55}}&\textcolor{darkgreen}{\ding{51}}\\
		
		RepoCoder \citep{RepoCoder}&\textcolor{red}{\ding{55}}&\textcolor{red}{\ding{55}}&\textcolor{darkgreen}{\ding{51}}&\textcolor{red}{\ding{55}}\\
		
		RLCoder \citep{RLCoder}&\textcolor{red}{\ding{55}}&\textcolor{red}{\ding{55}}&\textcolor{darkgreen}{\ding{51}}&\textcolor{red}{\ding{55}}\\
		
		CoCoGen \citep{CoCoGen}&\textcolor{red}{\ding{55}}&\textcolor{darkgreen}{\ding{51}}&\textcolor{red}{\ding{55}}&\textcolor{red}{\ding{55}}\\\midrule
		
		\rowcolor{mygray}\textbf{TICoder}&\textcolor{darkgreen}{\ding{51}}&\textcolor{darkgreen}{\ding{51}}&\textcolor{darkgreen}{\ding{51}}&\textcolor{darkgreen}{\ding{51}}\\
		\bottomrule
	\end{tabular}
	\vspace{-9pt}
\end{table}

\section{Related Work}\label{Related Work}

\subsection{Repository-level Code Generation}

Existing approaches adopt RAG techniques on repository-level code generation tasks, due to complex dependencies and limited context windows \citep{REPOFORMER, AllianceCoder, RepoHyper, RepoScope, RepoCoder}. 

 
 Several approaches focus on selective retrieval. RepoFormer~\citep{REPOFORMER} introduces a mechanism to decide when retrieval is necessary during generation. 
 Probing-RAG~\citep{ProbingRAG} analyzes model hidden states to adaptively decide retrieval timing. 
 Some works enhance retrieval by modeling repository structures as graphs. CoCoMIC~\citep{CoCoMIC} and RepoHyper~\citep{RepoHyper} construct method-level graphs to capture dependencies between functions. GraphCoder~\citep{GraphCoder} further incorporates statement-level structures to better model fine-grained semantics. RepoGraph~\citep{RepoGraph} represents repositories as graph structures for improved code understanding. 
RepoScope~\citep{RepoScope} introduces call chain-aware multi-view context modeling for repository-level code generation.
Some works adopt iterative frameworks to refine generation. RepoCoder~\citep{RepoCoder} iteratively retrieves relevant code fragments and updates the generated code. De-Hallucinator~\citep{DeHallucinator} iteratively verifies and corrects generated code to reduce hallucinations. 


More recently, planning-based methods have been proposed to address the limitation that directly retrieving similar code snippets may fail when no closely matching implementation exists. CodePlan~\citep{CodePlan} formulates repository-level code generation as a planning problem by generating multi-step plan chains. AllianceCoder~\citep{AllianceCoder} decomposes developer requirements into implementation steps and retrieves relevant APIs for each step. \citet{self-planning} guided LLMs to incrementally generate code via intermediate planning. 

\begin{figure*}[t]
	\includegraphics[width=\linewidth]{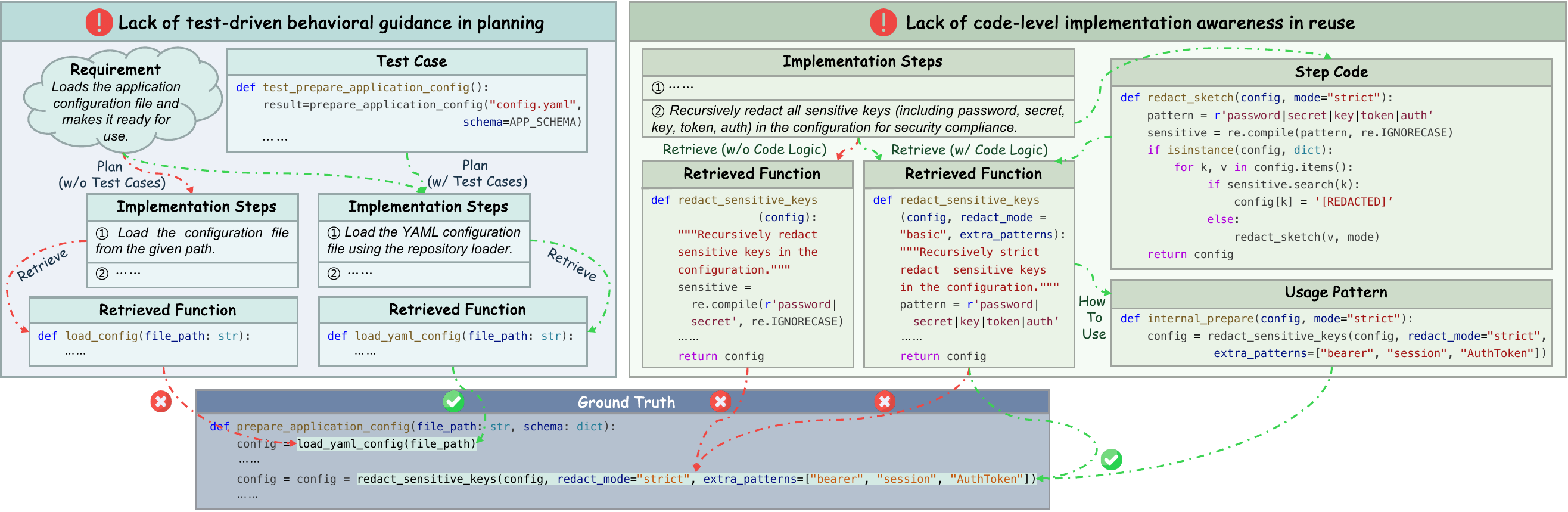}
	\vspace{-7pt}
	\caption{A motivating example shows the limitations of prior works: lack of test-driven behavioral guidance in planning and lack of code-level implementation awareness in reuse.}
	\label{fig_motivation}
	\vspace{-7pt}
\end{figure*}

Overall, existing approaches have improved repository-level code generation. However, generating precise planning and identifying potential callee functions for reuse remains a challenge. To address these limitations, we propose TICoder, which introduces a test-driven iterative planning mechanism to improve planning quality and an implementation-aware reuse strategy to enhance both retrieval and usage of repository functions.

\subsection{Test-Driven Code Generation}

Incorporating test cases into code generation has emerged as an effective paradigm for improving generation quality and reliability. Existing approaches primarily leverage test cases in three ways.
Some studies incorporate test cases directly into prompts to provide additional behavioral specifications. For example, \citet{TDDLLMCG} augmented problem descriptions with test cases to guide LLMs toward generating code that satisfies expected behaviors. LLM4TDD~\citep{LLM4TDD} further adopts a test-driven development paradigm, where LLMs iteratively generate code conditioned on test cases.
Another line of work utilizes test execution feedback to refine generated code. CoCoGen~\citep{CoCoGen} leverages compiler feedback and static analysis to detect inconsistencies between generated code and project-specific constraints. Similarly, \citet{TDMTCG} incorporated test execution signals during training, enabling models to distinguish between correct and incorrect code.
Reflexion~\citep{Reflexion} enables agents to iteratively improve code generation by learning from past failures, highlighting the importance of feedback-driven refinement in software development tasks.

Despite their effectiveness, existing approaches primarily apply test cases in the code generation or post-generation stages. The potential of leveraging test cases to guide the planning process, particularly for improving the quality of intermediate implementation steps, remains largely unexplored. This limitation motivates our work, which incorporates test cases into the planning stage through a test-driven iterative planning mechanism.

\section{Motivating Example}
To illustrate the challenges faced by existing planning-based repository-level code generation approaches \citep{AllianceCoder, CodePlan, self-planning}, we present a motivating example in Fig.~\ref{fig_motivation}. The task is to implement the function ``\emph{prepare\_application\_config}'', which loads the application configuration file and prepares it for use.

Given the requirement, existing approaches typically generate implementation steps directly from the natural language description, such as \emph{``load the configuration file from the given path''}. Although this step is partially correct, it fails to capture critical behavioral details required by the task, which consequently leads to inaccurate retrieval of callee functions. In contrast, introducing test cases provides explicit behavioral specifications for generating implementation steps. For example, test cases clarify that the configuration file is in YAML format. Guided by test-driven signals, the refined implementation step becomes \emph{``load the YAML configuration file using the repository loader''}, which is more precise and better aligned with repository-specific implementations. As a result, the refined plan enables accurate retrieval of relevant functions, such as \emph{``load\_yaml\_config''}. This example demonstrates the importance of test cases in producing more accurate implementation steps.

Even with correct implementation steps, existing approaches still face challenges in effectively retrieving and reusing relevant callee functions. As shown in Fig.~\ref{fig_motivation}, the task requires recursively redacting sensitive keys (e.g., \emph{``password''}, \emph{``token''}) in the configuration. When relying solely on semantic similarity between the implementation step and the function description, the retrieved function corresponds to a non-strict redaction setting, which is insufficient for handling strict sensitive key requirements. By further generating code representations of implementation steps and incorporating code implementation-level similarity, it becomes possible to retrieve the correct function that supports strict redaction.

However, correctly invoking this function depends on specific implementation details, such as parameter configurations (e.g., \emph{``redact\_mode''}). Providing only the retrieved function is insufficient for LLMs to learn how to properly use it. In contrast, examining usage patterns in the repository reveals how such functions are actually invoked in practice (e.g., via \emph{``internal\_prepare''}), which provides essential guidance for correct reuse. Without such usage patterns, even correctly retrieved functions may not be properly utilized in generated code.

\begin{figure*}[t]
	\includegraphics[width=\linewidth]{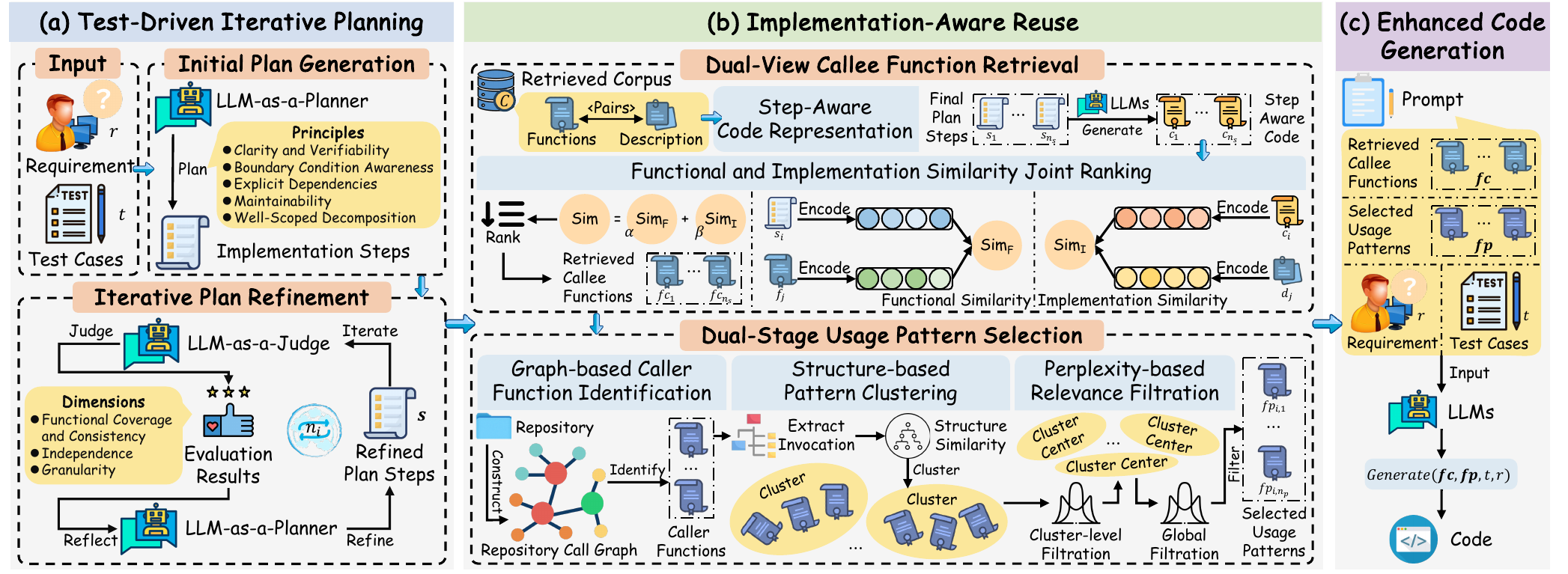}
	\vspace{-8pt}
	\caption{Overview of TICoder.}
	\label{fig_overflow}
	\vspace{-8pt}
\end{figure*}

Moreover, functions in repositories are often associated with multiple usage patterns. Selecting representative usage patterns is crucial to provide informative examples while avoiding excessive noise and token redundancy. Therefore, we further propose a dual-stage usage pattern selection strategy to identify representative usage patterns. The details are described in Section~\ref{pattern}.

\section{Approach}\label{Approach}

\subsection{Overview}
We propose TICoder, a novel repository-level code generation framework with test-driven iterative planning and implementation-aware code reuse. 
As shown in Fig. \ref{fig_overflow}, TICoder comprises three stages, including test-driven iterative planning, implementation-aware reuse, and enhanced code generation.

\subsection{Problem Formulation}
Based on the developer's requirement $r$ and test cases $t$, TICoder first generates implementation steps $\mathbf{s} = \{s_{1}, \cdots, s_{{n_{s}}}\}$, where $n_{s}$ is the number of generated implementation steps, through a test-driven iterative planning mechanism ($\S$ \ref{Plan}). 

For each step $s_i$, TICoder retrieves a set of potential callee functions $fc_{{i}}$ from the corpus $\mathbf{C} = \{(f_1, d_1),\cdots,(f_{n_{c}}, d_{n_{c}})\}$. Here, $f_{j}$ and $d_{j}$  represent a function, and its corresponding description in the repository, respectively, and $n_c$ denotes the number of functions in the corpus. This process yields a final set of retrieved callee functions $\mathbf{fc}$. Subsequently, we identify the caller functions for each $fc_{{i}}$ and retain the most relevant usage patterns, forming a set $\mathbf{fp}_{{i}} = \{fp_{{i,1}},\cdots,fp_{{i,n_{p}}}\}$, where $n_{p}$ represents the number of selected usage patterns for each  $fc_{{i}}$. The complete collection of selected patterns across all implementation steps constitutes the final usage patterns set $\mathbf{fp}$. The retrieval and selection processes are guided by an implementation-aware repository grounding strategy ($\S$ \ref{Ground}).

Finally, the enhanced code generation can be formulated as $Code=Generate(\mathbf{fc}, \mathbf{fp}, t, r)$, where $Generate(\cdot)$ denotes utilizing LLMs to generate the desired code ($\S$ \ref{Generate}).

Specifically, the relevance for retrieving potential callee functions is measured by computing the cosine similarity between vector representations.

For selecting pertinent usage patterns, relevance is assessed based on perplexity (PPL).  Specifically, for each candidate context $c$, we define $\operatorname{PPL}(r\mid c)$ as the conditional perplexity of $r$ given $c$, where lower values indicate a higher likelihood of $r$:
\begin{equation}\label{PPL}
	\operatorname{PPL}(r \mid c) = \exp\left(-\frac{1}{n_r} \sum_{i=1}^{n_r} \log P(r_i \mid r_{< i}, c)\right),
\end{equation}
\noindent where $P$ denotes the model's next-token prediction probability, $n_r$ is the sequence length of $r$, and $r_{< i}$ is the sequence of preceding tokens before $r_i$. A lower PPL score suggests that context $c$ enables the model to predict $r$ more accurately. Therefore, by retaining usage patterns associated with the lower PPL, we ensure that the core information necessary for enhancing the code generation process is preserved.

\subsection{Test-Driven Iterative Planning}\label{Plan}

While requirement descriptions provided to developers typically specify the core functionality, they often omit detailed descriptions of edge cases, making it challenging for LLMs to comprehensively identify and handle boundary scenarios during planning. In contrast, test cases provide rich information about edge conditions and expected behaviors, which can effectively complement requirement descriptions and reduce ambiguity. 

Therefore, we introduce test-driven iterative planning to generate high-quality implementation steps for the given requirement. This process consists of two stages, initial plan generation and iterative plan refinement via judge-and-reflection, involving two roles, an \emph{LLM-as-a-planner} for generating implementation steps and an \emph{LLM-as-a-judge} for evaluating and refining them.

\subsubsection{Initial Plan Generation.}

We first employ the \emph{LLM-as-a-planner} to generate an initial sequence of implementation steps based on both the requirement description and the associated test cases. To guide this process, we adopt five principles inspired by \emph{IEEE Software Requirement Specifications (SRS)} and prior work \citep{ClarifyGPT}:

\begin{itemize}
  \item \textbf{Clarity and Verifiability.} Each implementation step should be clearly defined, unambiguous, and verifiable with respect to the requirement and test cases.

  \item \textbf{Boundary Condition Awareness.} Implementation steps should explicitly account for edge cases (e.g., invalid inputs, null values) and expected handling behaviors.

  \item \textbf{Explicit Dependencies.} Dependencies on external functions, modules, or libraries should be clearly identified to support accurate downstream retrieval.

  \item \textbf{Maintainability.} Implementation steps should be structured to facilitate future modifications and extensions.

  \item \textbf{Well-Scoped Decomposition.} Each implementation step should focus on a well-defined functionality without introducing unnecessary side effects.
\end{itemize}

Guided by these principles, the \emph{LLM-as-a-planner} generates an initial sequence of implementation steps, with the number of steps adaptively determined but bounded by a predefined limit $n_s$.

\subsubsection{Iterative Plan Refinement via Judge-and-Reflection.}

To further improve the quality of implementation steps, we employ an iterative refinement process based on a judge-and-reflection mechanism. In each iteration, the \emph{LLM-as-a-judge} evaluates the current implementation steps and provides structured feedback, which is then used by the \emph{LLM-as-a-planner} to refine the plan.

The evaluation is conducted along three key dimensions, inspired by \emph{SRS} principles and prior works \citep{SodaCoder, ChatCoder}:

\begin{itemize}
  \item \textbf{Functional Coverage and Consistency.} Evaluates whether the implementation steps collectively cover all required functionalities and remain consistent with the behaviors specified by the test cases.

  \item \textbf{Independence.} Assesses the degree to which individual implementation steps are modular and can be implemented or modified with minimal dependency on others.

  \item \textbf{Granularity.} Measures whether the level of decomposition is appropriate, avoiding both overly coarse steps that lack actionable detail and overly fine steps that introduce unnecessary complexity.
\end{itemize}

Each dimension is scored on a scale from 0 to 100, along with textual justifications. If all scores exceed a predefined threshold, the current implementation steps are considered satisfactory. Otherwise, the evaluation feedback, together with the requirement description, test cases, and current implementation steps, is fed back to the \emph{LLM-as-a-planner} for further refinement.

The process iterates until the refined steps meet the quality criteria or a maximum number of iterations $n_i$ is reached.

\subsection{Implementation-Aware Reuse}\label{Ground}
In this stage, TICoder retrieves potential callee functions from the repository and identifies representative usage patterns to support effective reuse. Specifically, we adopt a dual-view retrieval strategy to improve retrieval precision and a dual-stage selection strategy to identify high-quality usage patterns, enabling more accurate and practical reuse of repository functions.

\subsubsection{Dual-View Callee Function Retrieval}
We introduce dual-view similarity that considers both functional similarity and implementation similarity to retrieve potential callee functions from the repository. Functional similarity captures the semantic relevance between implementation steps and function descriptions, while implementation similarity captures the alignment between generated code representations and function implementations. By jointly considering both views, TICoder can retrieve functions that are not only semantically relevant but also exhibit similar implementation logic. This design mitigates the limitation of purely semantic retrieval by incorporating implementation-level signals.

\textbf{Step-Aware Code Representation.} 
Given the generated implementation steps $\mathbf{s}$, we construct code-level representations for each step. Specifically, for each step $s_i$, we employ LLMs to produce a corresponding self-contained code snippet $c_i$, which serves as the step-aware code representation of $s_i$.

\textbf{Functional and Implementation Similarity Joint Ranking.}
For each implementation step $s_i$ with its code representation $c_i$, and each candidate function $f_j$ with description $d_j$ in the repository corpus $\mathbf{C}$, we construct embedding representations $\mathbf{E}_{s_i}$, $\mathbf{E}_{c_i}$, $\mathbf{E}_{f_j}$, and $\mathbf{E}_{d_j}$.  
We define the functional similarity $Sim_F(i, j)$ as the cosine similarity between $\mathbf{E}_{s_{i}}$ and $\mathbf{E}_{d_{j}}$, as well as the implementation similarity $Sim_I(i, j)$ as the cosine similarity between $\mathbf{E}_{c_{i}}$ and $\mathbf{E}_{f_{j}}$.

We compute a joint similarity score $Sim(i, j)$ between $s_{i}$ and each pair $(f_{j}, d_{j})$ of function and description using a weighted sum:
\begin{equation}
	Sim(i, j) = \alpha\times Sim_F(i, j)+\beta\times Sim_I(i, j),
\end{equation}
\noindent where $\alpha$ and $\beta$ represent the weight of $Sim_F$ and $Sim_I$, respectively, as well as $\alpha + \beta = 1$.

We rank all candidate functions $f_j \in \mathbf{C}$ based on $Sim(i, j)$ in descending order and select the top-ranked function as the most relevant callee function $fc_i$ for step $s_i$. This results in a set of retrieved callee functions $\mathbf{fc} = \{fc_1, \ldots, fc_{n_s}\}$.

\begin{algorithm}[t]
    \small
    \caption{The algorithm for dual-stage representative usage patterns selection.}
    \label{algorithm_prune}
    \begin{algorithmic}[1]
        \renewcommand{\algorithmicrequire}{\textbf{Input:}}
		    \renewcommand{\algorithmicensure}{\textbf{Output:}}
        \REQUIRE $\mathbf{fc}$: The set of all identified callee functions. $RCG$: The repository call graph. $n_p$: The maximum number of usage patterns to select per callee function.
		\ENSURE $\mathbf{fp}$: The dictionary mapping each callee function to its selected usage patterns.
        \STATE $\mathbf{fp} \gets \emptyset$ 
        \FOR{each callee function $c \in \mathbf{fc}$}
            \STATE $callers \gets \{f \mid (f \to c) \in RCG\}$ 
            \STATE $clusters \gets \emptyset$ 
        \FOR{each caller function $f \in callers$}
            \STATE $callees_f \gets \{g \mid (f \to g) \in RCG\}$ 
            \STATE $key \gets \text{Sort}(callees_f)$ 
            \IF{$key \notin clusters$}
                \STATE $clusters[key] \gets \emptyset$
            \ENDIF
            \STATE $clusters[key].\text{add}(f)$
        \ENDFOR
        \STATE $cluster\_centers \gets \emptyset$
        \FOR{each $key \in clusters$}
            \STATE $cluster \gets clusters[key]$
            \STATE $center \gets \arg\min_{f \in cluster} \text{PPL}(f)$ 
            \STATE $centers.\text{add}(center)$
        \ENDFOR
        
        \STATE $\text{sort}(centers)$ by $\text{PPL}$ in ascending order
        \STATE $selected \gets \text{first } \min(n_p, |centers|) \text{ of } centers$
        \STATE $\mathbf{fp}[c] \gets selected$
    \ENDFOR
    \RETURN $\mathbf{fp}$
    \end{algorithmic}
\end{algorithm}

\subsubsection{Dual-Stage Usage Pattern Selection} \label{pattern}
To further enable effective reuse, we select representative usage patterns for each retrieved callee function. Since a callee function may be invoked in diverse contexts, we propose a dual-stage selection strategy to filter and retain high-quality usage patterns. This design allows TICoder to capture how functions are used in practice, reducing noise and conserving computational resources, rather than relying solely on retrieved code snippets.

\textbf{Graph-based Caller Function Identification.} 
We employ an open-source tool LSP to construct the Repository Call Graph (RCG), which captures function invocation relationships within the repository. For each retrieved callee function $fc_i$, we identify all caller functions in the RCG as candidate usage patterns.

\textbf{Structure-based Pattern Clustering.}
We cluster candidate caller functions based on structural similarity. Specifically, for each caller function, we extract its set of invoked functions from the RCG. Caller functions that invoke identical sets of functions are considered structurally equivalent and grouped into the same cluster.

\textbf{Perplexity-based Relevance Filtration.}
We conduct a perplexity-based relevance filtration in two steps, cluster-level filtering and global filtering.  Each cluster contains caller functions with the same structure, which generally enable LLMs to learn similar usage patterns. Therefore, within each cluster, we retain the most relevant caller function as the cluster center based on PPL (Equation \ref{PPL}). For all cluster centers, we perform global filtering by ranking them in ascending order of relevance based on PPL to form a candidate list of usage patterns. Finally, for each retrieved callee function $fc_i$, we select $top_{n_{p}}$ usage patterns, $\mathbf{fp}_i = \{fp_{i,1},\cdots,fp_{i, n_p}\}$, from the candidate list to serve as example instructions. 

Specifically, we introduce the algorithm for dual-stage usage patterns selection in Algorithm \ref{algorithm_prune}. In lines 4-6, for each callee function, we identify all functions that call it as candidate usage patterns. In lines 5-11, we cluster usage patterns based on structural similarity. In lines 14-18, we select the most relevant function as the cluster center based on PPL within each cluster. In lines 19-21, we utilize global filtering to sort all cluster centers in ascending order of PPL and select the top as representative usage patterns.

\subsection{Enhanced Code Generation}\label{Generate}
We concatenate the retrieved callee functions $\mathbf{fc}$ along with selected usage patterns $\mathbf{fp}$, test cases $t$, and the original requirement $r$ as the input prompt to LLMs for enhanced code generation. 


\begin{table*}[t]
  \footnotesize
  \caption{Overall performance comparison across studied benchmarks and backbone models. Results are shown in percentage (\%). The best results are highlighted in bold, and the best-performing baseline is underlined.}
	\label{table_overall}
		\vspace{-5pt}
	\setlength{\tabcolsep}{2.8pt}
	\centering
	\renewcommand{\arraystretch}{1}
	\begin{tabular}{l|lll|lll|lll}
		\toprule
		\multirow{2.5}{*}{\textbf{Approaches}} & \multicolumn{3}{c|}{\textbf{GPT-4o-mini}}           & \multicolumn{3}{c}{\textbf{DeepSeek-V3}}  & \multicolumn{3}{c}{\textbf{Qwen2.5-Coder-7B}}      \\ \cmidrule(r){2-4} \cmidrule(r){5-7} \cmidrule(r){8-10}
		& \textbf{Pass@1} & \textbf{Pass@3} & \textbf{Pass@5} & \textbf{Pass@1} & \textbf{Pass@3} & \textbf{Pass@5} & \textbf{Pass@1} & \textbf{Pass@3} & \textbf{Pass@5} \\\midrule
        \rowcolor{mygray}\multicolumn{10}{c}{\emph{\textbf{CoderEval (Python)}}}\\\midrule
		SimpleRAG&    33.91 & 40.55    &42.66   & 41.30	 &47.06	&48.69&     26.96 & 	34.84	 & 38.57\\
		RepoCoder& 30.87&	38.39&	40.49 & 32.61& 	38.93& 	41.12& 31.74	 & 39.07 & 	42.28 \\
		A$^3$Codgen& 23.48	&31.58&	34.43& 39.57& 44.49& 	46.93& 23.48 & 	28.76 & 	30.95 \\
		AllianceCoder&\underline{48.70}	&\underline{51.20} &	52.36&\underline{52.17}& \underline{54.10}& \underline{54.63}  &     \underline{40.43} & 	\underline{45.64}	 & \underline{47.41}\\
		RLCoder& 31.30	&39.22&	43.04 & 33.48& 40.20& 43.29 &  29.57 & 	36.27 & 	39.14\\
		RepoScope& 37.61&	49.13&	\underline{52.57}& 43.36& 51.19& 	54.30 & 22.17 & 	34.41 	 & 38.80\\\midrule
		\rowcolor{myblue}\textbf{TICoder} & \textbf{49.57} ($\uparrow$1.79\%)  &   \textbf{56.27}	 ($\uparrow$9.90\%)& \textbf{58.61} ($\uparrow$11.49\%)& \textbf{56.09}	 ($\uparrow$7.90\%)& \textbf{63.78}  ($\uparrow$17.89\%)& \textbf{66.04} ($\uparrow$20.89\%)& \textbf{41.30} ($\uparrow$2.15\%)	 &  \textbf{46.62} ($\uparrow$2.15\%)	 &  \textbf{49.90} ($\uparrow$5.25\%)\\ \midrule
        
        \rowcolor{mygray}\multicolumn{10}{c}{\emph{\textbf{CoderEval (Java)}}}\\\midrule
        SimpleRAG &  40.87&	43.59	&45.16&45.46&	46.33	&46.64 & \underline{38.70} 	&\underline{41.60} &	\underline{43.56}\\
		RepoCoder  & 35.65&	39.17&	40.36 & 36.52&	40.26	&40.96&  20.87&	26.98	&30.73  \\
		A$^3$Codgen& 38.70	&44.18&	45.87 & 51.74	&54.55&	55.50 & 23.04&	24.18&	27.78  \\
		AllianceCoder& \underline{43.91} &	44.31	&44.53&  \underline{53.04} &	54.33	&55.10 &  35.65	&40.70	&42.37 \\
		RLCoder& 35.22	&\underline{46.48} 	&\underline{49.88}   & 41.74&	52.56&	55.41 &   32.17	&38.95&	44.06\\
		RepoScope &  37.39&	43.54	&45.74 & 53.91&	\underline{55.93} 	&\underline{57.43}  & 27.39	&37.29	&41.27 \\\midrule
		\rowcolor{myblue}\textbf{TICoder}  & \textbf{ 51.30}  ($\uparrow$16.83\%)	&\textbf{55.42} ($\uparrow$19.23\%) &	\textbf{57.20}  ($\uparrow$14.68\%) & \textbf{55.65}  ($\uparrow$4.92\%)&	\textbf{57.75}  ($\uparrow$3.25\%)&	\textbf{58.37} ($\uparrow$1.64\%)&  \textbf{44.35}  ($\uparrow$14.60\%)&\textbf{50.92}  ($\uparrow$22.40\%)&	\textbf{53.44} ($\uparrow$22.68\%)\\\midrule

        \rowcolor{mygray}\multicolumn{10}{c}{\emph{\textbf{DevEval}}}\\\midrule
        SimpleRAG  &  24.05	& 28.42	& 30.20 & 31.95	& 36.63	& 37.96&  15.56  & 20.99 &  23.49\\
		RepoCoder  & 15.01& 	17.48	& 18.52& 20.60	& 25.34	& 27.20 & 12.00	& 16.12	& 17.93  \\
		A$^3$Codgen & 21.81	& 25.72	& 27.49&  36.49	& \underline{41.20}	& \underline{42.95}  & 17.04& 	21.20	& 23.49\\
		AllianceCoder& \underline{25.26}	& 28.06	& 29.01& \underline{36.54}	& 38.64	& 39.48  & \underline{21.09}& 	\underline{24.52}& 	\underline{26.11}\\
		RLCoder & 18.58	& 20.60	& 21.94& 22.08	& 27.43	& 29.29 &  14.58	& 17.85	& 19.74  \\
		RepoScope &23.07	& \underline{31.70}	& \underline{34.69} & 29.59	& 39.74	& 42.87  &  16.71	& 19.73	& 22.83  \\\midrule
		\rowcolor{myblue}\textbf{TICoder} & \textbf{ 31.01}	 ($\uparrow$22.76\%)& \textbf{35.78} ($\uparrow$12.87\%)	& \textbf{37.64}  ($\uparrow$8.50\%)& \textbf{ 43.29}	 ($\uparrow$18.47\%)& \textbf{47.08} ($\uparrow$14.27\%)	& \textbf{48.52}   ($\uparrow$12.97\%)& \textbf{21.92}	 ($\uparrow$3.94\%)& \textbf{26.47} ($\uparrow$7.95\%)	& \textbf{28.67} ($\uparrow$9.80\%)\\

		\bottomrule         
	\end{tabular}
		\vspace{-5pt}
\end{table*}

\section{Experimental Setup}
\subsection{Research Questions (RQs)}
In this work, we aim to answer the following research questions:
\begin{itemize}
  \item \textbf{RQ1: Overall Performance.} How does TICoder perform on repository-level code generation compared to baselines?

  \item \textbf{RQ2: Ablation Study.} To what extent do the key components and strategies of TICoder contribute to its overall performance?

  \item \textbf{RQ3: Impact of Planning Iterations.} How does the number of planning iterations affect the performance of TICoder?

  \item \textbf{RQ4: Impact of Similarity Weights.} How do different weights assigned to functional and implementation similarity affect the performance of TICoder?

  \item \textbf{RQ5: Impact of Usage Patterns.} How does the number of selected usage patterns influence the performance of TICoder?
\end{itemize}

\subsection{Datasets}
Our experiments were conducted on two widely-used repository-level code generation benchmarks, CoderEval \citep{CoderEval} and DevEval \citep{DevEval}, which are both used to evaluate code generation performance for the developer's requirement. CoderEval consists of 230 Python tasks from 43 projects and 230 Java tasks from 10 projects. DevEval consists of 1825 Python tasks from 115 real-world open-source projects on GitHub.
Each task in CoderEval and DevEval contains a human-written requirement and comprehensive test cases to verify functional correctness.

\subsection{Baselines and LLMs Evaluated}\label{baselines}
We evaluated the retrieval-augmented code generation approaches as baselines. 

\begin{itemize}

	\item \textbf{SimpleRAG} retrieves relevant code snippets based on functional similarity and appends them to the prompt for LLMs.
	
	\item \textbf{RepoCoder} \citep{RepoCoder} is a retrieval-augmented framework in an iterative retrieval-generation pipeline.
	
	\item \textbf{A$^3$Codgen} \citep{A3CodGen} utilizes local-aware, global-aware, and third-party-library information to augment code generation.
	
	\item \textbf{AllianceCoder} \citep{AllianceCoder} aims to retrieve and append potential APIs from the repository to the user query before being fed into LLMs.
	
	\item \textbf{RLCoder} \citep{RLCoder} is a reinforcement learning-based framework for optimizing the retriever.
	
	\item \textbf{RepoScope} \citep{RepoScope} leverages call chain-aware multi-view context for repository-level code generation.
	
\end{itemize}
We evaluated baselines and our approach on three mainstream LLMs, GPT-4o-mini \citep{GPT}, DeepSeek-V3 \citep{DeepSeek}, and Qwen2.5-Coder-7B \citep{Qwen2.5-Coder}.

\subsection{Metrics}
Following previous studies \citep{DevEval, CoderEval, RepoCoder, AllianceCoder}, we evaluated the functional correctness of generated code by executing test cases and computing the Pass@k.
\begin{equation}
	Pass@k := \underset{r}{\mathbb{E}}\left[1-\frac{\binom{n_g-c}{k}}{\binom{n_g}{k}}\right],
\end{equation}
\noindent where $n_g$ (with $n_g\geq k$) represents the number of generated code for each requirement $r$ and $c$ (with $c\le n_g$) represents the number of correct code that passes test cases.


\subsection{Implementation Details}
For TICoder, in the planning stage, we set the maximum number of implementation steps $n_s$ to 5, considering computational cost and efficiency. The selection of planner and judge LLMs, as well as the number of iterations $n_i$, are discussed in Section~\ref{experiment_iteration}. In the retrieval module, we use GPT-4o-mini to generate code representations corresponding to the refined implementation steps. The predefined threshold is set to 90 for the LLM-as-a-judge. We adopt OpenAI’s \emph{text-embedding-3-small} model to encode implementation steps, their corresponding generated code, as well as repository functions and their descriptions into vector representations for similarity computation. The weight settings for functional similarity $\alpha$ and implementation similarity $\beta$ are detailed in Section~\ref{experiment_weight}. In the usage pattern selection module, the number of selected usage patterns $n_p$ for each retrieved callee function is discussed in Section~\ref{experiment_usage}. For baseline methods, we follow the experimental settings described in their original papers. In the final code generation stage, as shown in Section~\ref{baselines}, we adopt three representative LLMs as backbone models for evaluation. Specifically, for GPT-4o-mini, we use its latest model snapshot \emph{gpt-4o-mini-2024-07-18}. For DeepSeek-V3, we use its latest version \emph{DeepSeek-V3-0324}. For Qwen2.5-Coder, we use its base model with 7B parameters. Additionally, for GPT-4o-mini and DeepSeek-V3, we access them via their official APIs, while for Qwen2.5-Coder, we use the open-source base model available on Hugging Face.

\section{Experimental Results and Analysis}

\subsection{RQ1: Overall Performance}
The overall performance comparison on CoderEval (Python and Java) and DevEval across the evaluated LLMs, including GPT-4o-mini, DeepSeek-V3, and Qwen2.5-Coder-7B, as the generation backbone models, is presented in Table~\ref{table_overall}. Specifically, in this RQ, we apply GPT-4o-mini for the \emph{LLM-as-a-planner} and \emph{LLM-as-a-judge}. Overall, TICoder consistently outperforms all baselines with an increase of 11.25\% across different benchmarks, programming languages, and backbone models, demonstrating its effectiveness in repository-level code generation. To further validate the reliability of these improvements, we conduct statistical significance testing by comparing TICoder with the strongest baseline, AllianceCoder, across all settings. A two-tailed t-test yields $p < 0.05$, confirming that the observed performance gains are statistically significant rather than due to random variation. Notably, the largest improvement is observed on DevEval with the GPT-4o-mini backbone, where TICoder surpasses AllianceCoder by 22.76\%, indicating its strong capability in realistic repository-level scenarios.

A closer examination of the baselines reveals that RepoCoder and RLCoder often underperform, in some cases even falling behind SimpleRAG. We attribute this to their reliance on sliding-window and split-aggregate strategies for code segmentation, which may disrupt the structural integrity of code and break meaningful functional boundaries. In contrast, SimpleRAG treats functions as the basic retrieval unit, preserving the coherence of code logic and thus achieving more stable performance. Among all baselines, AllianceCoder consistently achieves the best results. Its advantage lies in introducing a planning mechanism that decomposes developer requirements into implementation steps, thereby improving the retrieval of relevant repository components. This observation highlights the importance of planning in repository-level code generation. Building upon this insight, TICoder further improves performance compared to AllianceCoder, highlighting the benefits of test-driven iterative planning and implementation-aware code reuse.\\
\begin{mdframed}[roundcorner=30pt, backgroundcolor=gray!30]
\textbf{Answer to RQ1:} TICoder consistently outperforms all baselines across evaluated benchmarks and LLMs, achieving a relative improvement of up to 11.25\%. The improvements are statistically significant ($p < 0.05$), highlighting the performance of TICoder on repository-level code generation.
\end{mdframed}

\subsection{RQ2: Ablation Study}
\begin{table}[t]
	\small
  \caption{Ablation study for TICoder on DevEval.}
	\label{table_ablations}
		\vspace{-5pt}
	\setlength{\tabcolsep}{3pt}
	\centering
	\renewcommand{\arraystretch}{1}
	\begin{tabular}{lcccc}
		\toprule
		\multirow{2.5}{*}{\textbf{Components}} & \multicolumn{4}{c}{\textbf{DevEval}} \\\cmidrule(r){2-3}\cmidrule(r){4-5}
		& \multicolumn{1}{c}{\textbf{Pass@1~}} & \multicolumn{1}{c}{\textbf{$\Delta$}} & \multicolumn{1}{c}{\textbf{\# Token}} & \multicolumn{1}{c}{\textbf{\textbf{$\Delta$}}} \\ \midrule
		\rowcolor{myblue}TICoder  &   31.01      &     -   &    2824    &    -     \\ \midrule
		\multicolumn{5}{l}{\textbf{\emph{Component 1: Planning}}}  \\ 
		~~~w/o planning & 28.88 &  \textcolor{red}{($\downarrow$6.87\%)} & 2730&   \textcolor{darkgreen}{($\downarrow$3.33\%)}    \\
		~~~w/o test cases   &      30.52   &     \textcolor{red}{($\downarrow$1.58\%)}     & 2921 &   \textcolor{red}{($\uparrow$3.43\%)}      \\
		~~~w/o iteration  &  30.74 &    \textcolor{red}{($\downarrow$0.87\%)}   & 2927 &  \textcolor{red}{($\uparrow$5.24\%)}  \\ \midrule
		\multicolumn{5}{l}{\textbf{\emph{Component 2: Retrieval}}}    \\ 
		~~~w/ $Sim_F$     & 30.25  &  \textcolor{red}{($\downarrow$2.45\%)}  & 2863 &  \textcolor{red}{($\uparrow$1.38\%)}  \\
		~~~w/ $Sim_I$   & 30.19 & \textcolor{red}{($\downarrow$2.64\%)} & 2851& \textcolor{red}{($\uparrow$0.96\%)} \\ \midrule
		\multicolumn{5}{l}{\textbf{\emph{Component 3: Selection}}}          \\ 
		~~~w/o patterns & 29.97& \textcolor{red}{($\downarrow$3.35\%)} & 1407  &  \textcolor{darkgreen}{($\downarrow$50.18\%)}     \\
		~~~w/o selection   & 30.90 &  \textcolor{red}{($\downarrow$0.35\%)}   &3523 &  \textcolor{red}{($\uparrow$24.82\%)}  \\
		~~~w/o clustering   &   30.58   &  \textcolor{red}{($\downarrow$1.39\%)} & 2848 &  \textcolor{red}{($\uparrow$0.85\%)}  \\ \midrule

    \multicolumn{5}{l}{\textbf{\emph{Component 4: Generation}}}  \\
    ~~~w/o test cases & 27.78 & \textcolor{red}{($\downarrow$10.42\%)} & 2792 &  \textcolor{darkgreen}{($\downarrow$1.13\%)} \\
    ~~~w/ implementation steps & 14.63 & \textcolor{red}{($\downarrow$52.82\%)} & 3214 &  \textcolor{red}{($\uparrow$13.81\%)} \\

		\bottomrule                             
	\end{tabular}
		\vspace{-5pt}
\end{table}

To evaluate the effectiveness of each key component and strategy in TICoder, we conduct an ablation study on the DevEval dataset using GPT-4o-mini as the backbone model, as shown in Table \ref{table_ablations}. Specifically, we first assess the contribution of the test-driven iterative planning component by sequentially removing planning, test cases, and iterative judge-and-reflection (w/o iteration), and observing the resulting changes in code generation performance. Then, we validate the effectiveness of the dual-view callee function retrieval component by constructing two variants: one using only functional similarity (w/ $Sim_F$) and the other using only implementation similarity (w/ $Sim_I$). Next, we evaluate the dual-stage usage pattern selection component by removing usage patterns (w/o patterns), removing the dual-stage selection (w/o selection), and removing the structure-based clustering (w/o clustering). Finally, we evaluate the enhanced code generation component by removing test cases and replacing the original requirement with generated implementation steps (w/ implementation steps).
As shown in Table \ref{table_ablations}, removing any component leads to performance degradation, demonstrating that all modules contribute positively to the final performance, while affecting token consumption differently.

Removing the planning module results in a decrease of 6.87\% in Pass@1, indicating that planning is critical for decomposing requirements. When removing test cases from planning, performance decreases by 1.58\%, suggesting that test cases help generate more concise and reliable implementation steps. Removing iterative refinement also leads to a decrease of 0.87\%, indicating that the judge-and-reflection process improves planning effectiveness.
Using only functional similarity or only implementation similarity results in performance drops of 2.45\% and 2.64\%, respectively. This demonstrates that both semantic relevance and implementation-level similarity are necessary for precise callee function retrieval. 
Removing usage patterns leads to a drop by 3.35\% in Pass@1, indicating that incorporating usage patterns improves performance. Removing the dual-stage selection module degrades performance by 0.35\% and increases token usage by 24.82\%, suggesting that dual-stage selection effectively filters redundant patterns and controls context size. Removing clustering causes a drop of 1.39\%, showing that structural grouping helps retain diverse and representative usage patterns.
Removing test cases in the generation stage results in a substantial performance drop of 10.42\%, highlighting the critical role of test cases in aligning generated code with expected behaviors. Notably, even when test cases are not incorporated during the generation stage, TICoder still consistently outperforms the strongest baseline. Additionally, replacing the original requirement with implementation steps causes a degradation by 52.82\%, indicating that preserving the original requirement is essential, and implementation steps alone are insufficient to guide final code generation. \\
\begin{mdframed}[roundcorner=30pt, backgroundcolor=gray!30]
\textbf{Answer to RQ2:} Each proposed key component and strategy in TICoder positively contributes to the overall performance. These results demonstrate that TICoder effectively integrates planning, retrieval, selection, and generation to achieve superior performance.
\end{mdframed}

\subsection{RQ3: Impact of Planning Iterations}\label{experiment_iteration}
\begin{table}[t]
		\small
    \caption{Results for the number of iterations on CoderEval (Python and Java) in Pass@1.}
	  \label{table_iterations}
	  	\vspace{-5pt}
		\centering
		\renewcommand{\arraystretch}{0.85}
		\setlength{\tabcolsep}{8pt}
	\begin{tabular}{c|c|cc}
		\toprule
		\multirow{2.5}{*}{\makecell{\textbf{LLMs}}} & \multirow{2.5}{*}{\textbf{\# Iterations}} & \multicolumn{2}{c}{\textbf{CoderEval}} \\ \cmidrule(r){3-4}
		&                                        & \textbf{Python}     & \textbf{Java}    \\ \midrule
		\multirow{5}{*}{GPT-4o-mini}   & 0                                      & 48.70             & 47.83         \\
		& 1  & 49.13   &     47.83             \\
		& 2  & 46.52   &      49.09            \\
		& 3  & 49.13   &     48.26             \\
		& 4  & \textbf{49.57}   &         \textbf{51.30}         \\\midrule

        \multirow{5}{*}{DeepSeek-V3}   & 0                                      & 45.65             & 48.70         \\
		& 1                                      &        47.83             &  49.13    \\
		& 2                                      &     44.78   &    \textbf{49.57}      \\
		& 3                                      &    45.22   &      45.96      \\
		& 4                                      &   \textbf{48.26}     &     48.26    \\
		\bottomrule
	\end{tabular}
	\vspace{-5pt}
\end{table}

In the planning stage, we employ an iterative refinement strategy based on judge-and-reflection. To evaluate its effectiveness, we vary the number of planning iterations from 0 to 4 and utilize two LLMs, GPT-4o-mini and DeepSeek-V3, to plan and judge, while fixing GPT-4o-mini as the backbone for code generation. 

As shown in Table~\ref{table_iterations}, iterative refinement generally improves performance compared to the non-iterative setting (0 iterations), demonstrating the effectiveness of the judge-and-reflection mechanism. 
However, the performance does not increase monotonically with the number of iterations. For instance, with GPT-4o-mini as the planner and judge, performance on Python drops at 2 iterations before recovering, and similar fluctuations are observed for DeepSeek-V3. This indicates that excessive iterations may introduce noise or over-refinement, leading to suboptimal planning results.
We further observe that the optimal number of iterations varies across LLMs and datasets. With GPT-4o-mini, the best performance is achieved at 4 iterations for both Python and Java. In contrast, with DeepSeek-V3, the optimal setting differs: 4 iterations for Python and 2 iterations for Java. Despite these differences, both planners benefit from iterative refinement and achieve competitive performance in their best configurations, demonstrating the robustness of the proposed planning framework.\\
\begin{mdframed}[roundcorner=30pt, backgroundcolor=gray!30]
\textbf{Answer to RQ3:} Increasing the number of planning iterations generally improves performance, as iterative judge-and-reflection enhances the quality of implementation steps. However, the improvement is not monotonic, and excessive iterations may introduce noise and degrade performance.
\end{mdframed}

\subsection{RQ4: Impact of Similarity Weights}\label{experiment_weight}
\begin{figure}[t]
	\centering
	\includegraphics[width=0.93\linewidth]{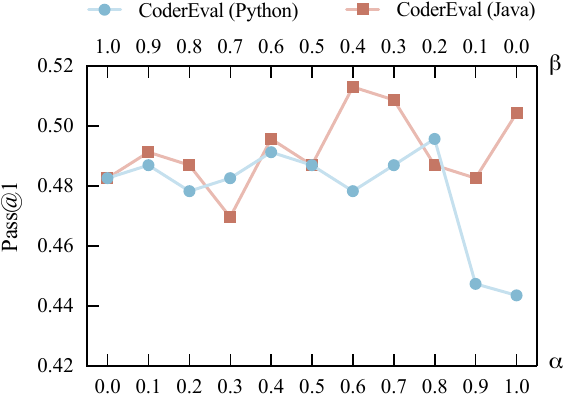}
		\vspace{-5pt}
	\caption{Performance changes with different weights for dual-view similarity calculation.}
	\label{fig_weight}
		\vspace{-5pt}
\end{figure}

In this section, we analyze the impact of weights between functional similarity and implementation similarity in the dual-view retrieval module. Specifically, we vary the weight $\alpha$ for $Sim_F$ from 0 to 1 with a step size of 0.1, while adjusting $\beta$ for $Sim_I$ accordingly such that $\alpha + \beta = 1$. Experiments are conducted on CoderEval (Python and Java) using GPT-4o-mini as the backbone.


As shown in Fig.~\ref{fig_weight}, the optimal performance is achieved at $\alpha = 0.8$ and $\beta = 0.2$ for CoderEval (Python), as well as $\alpha = 0.6$ and $\beta = 0.4$ for CoderEval (Java). These results indicate that while functional similarity is generally more important, incorporating implementation similarity further improves performance by capturing implementation-level execution logic.
Additionally, we observe that the optimal weight of implementation similarity is higher for Java than for Python. We attribute this difference to the statically-typed nature of Java, where implementation details provide stronger signals for functional behavior. In contrast, Python’s dynamic typing makes semantic descriptions relatively more informative than implementation structure.\\
\begin{mdframed}[roundcorner=30pt, backgroundcolor=gray!30]
\textbf{Answer to RQ4:} The performance of TICoder is sensitive to the weighting between functional and implementation similarity. Functional similarity generally plays a dominant role, but incorporating implementation similarity further improves the performance of code generation.
\end{mdframed}

\subsection{RQ5: Impact of Usage Patterns}\label{experiment_usage}

In this section, we analyze the impact of the number of selected usage patterns on the performance of TICoder. We conduct experiments on CoderEval (Python) using GPT-4o-mini as the backbone, with $n_p \in \{1,2,3\}$. The results are presented in Table~\ref{table_patterns}.

As shown in Table~\ref{table_patterns}, performance exhibits a non-monotonic trend with respect to $n_p$. Specifically, increasing $n_p$ from 1 to 2 improves Pass@1 by 5.56\%, indicating that incorporating multiple usage patterns provides richer contextual guidance for understanding how callee functions are used in practice. However, further increasing $n_p$ to 3 leads to a performance drop of 3.71\%, suggesting that excessive usage patterns may introduce redundant or noisy information for LLMs. Additionally, a consistent increase in token consumption as $n_p$ grows, highlighting a trade-off between contextual richness and efficiency. This finding further justifies the necessity of our dual-stage selection strategy for filtering informative and non-redundant usage patterns.\\
\begin{mdframed}[roundcorner=30pt, backgroundcolor=gray!30]
\textbf{Answer to RQ5:} The number of usage patterns has an impact on performance. 
Using too few patterns limits the model’s understanding of usage, while too many introduce noise and increase token consumption, degrading performance.
A moderate number of usage patterns (e.g., $n_p = 2$) yields the best performance by providing sufficient contextual guidance for reuse. 
\end{mdframed}

\begin{table}[t]
	\small
	\caption{Results for the number of selected usage patterns $n_p$ on CoderEval (Python) with GPT-4o-mini as the backbone.}
	\label{table_patterns}
	\vspace{-5pt}
	\centering
	\renewcommand{\arraystretch}{1}
	\setlength{\tabcolsep}{8pt}
	\begin{tabular}{c|cc}
		\toprule
		\textbf{$n_p$} & \textbf{Pass@1} & \textbf{Avg. \# Tokens}    \\ \midrule
		1 & 46.96& 2923\\
		2 & \textbf{49.57}& 3207\\
		3 & 45.22& 3298\\
		\bottomrule
	\end{tabular}
	\vspace{-5pt}
\end{table}

\section{Discussion}
\subsection{Case Study}

\begin{figure}[t]
	\centering
	\includegraphics[width=0.92\linewidth]{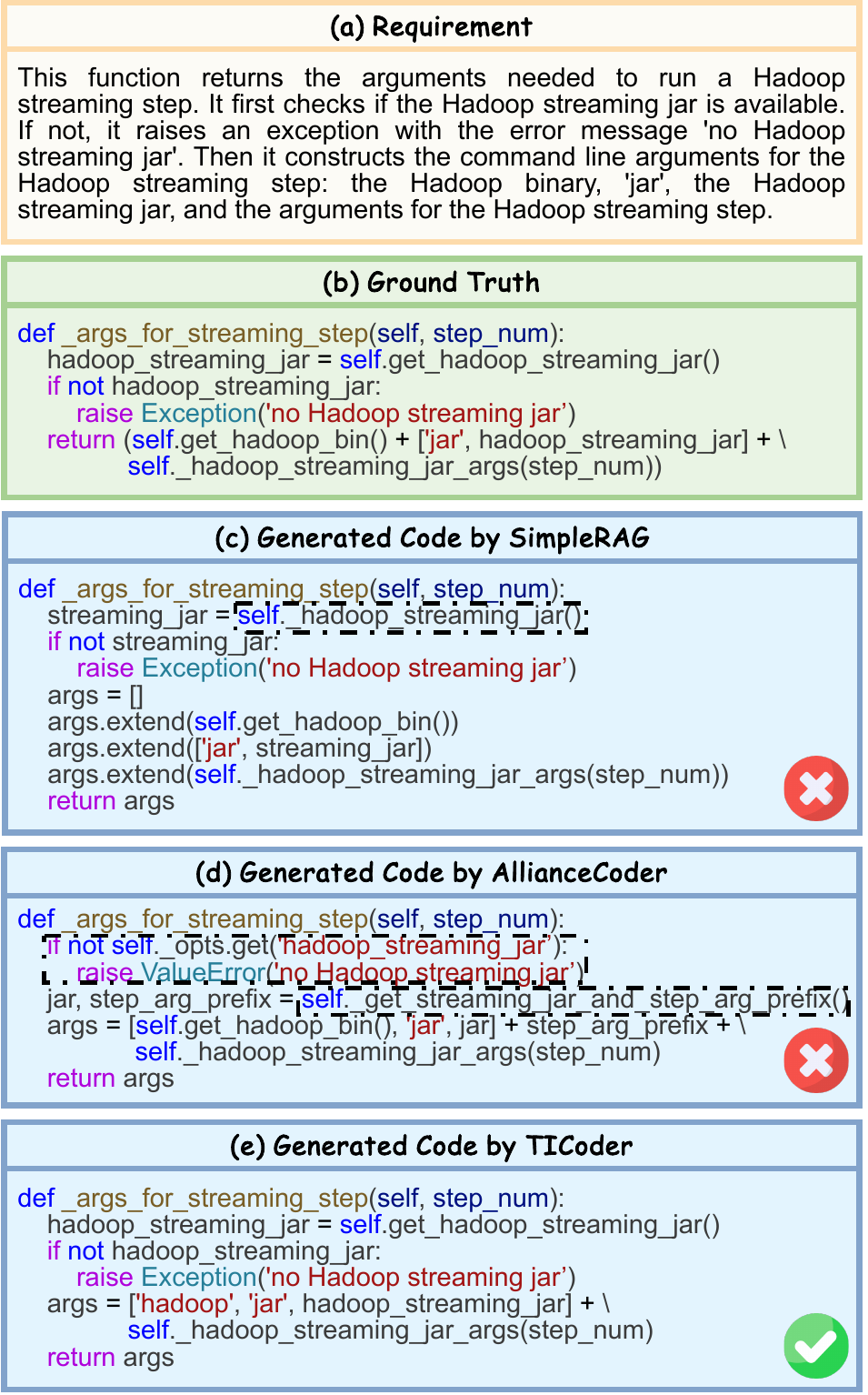}
	\vspace{-8pt}
	\caption{An example for a case study. The segments of code marked with black boxes indicate erroneous parts.}
	\label{fig_case}
	\vspace{-8pt}
\end{figure}

To intuitively demonstrate the effectiveness of TICoder, we present an example in Fig. \ref{fig_case}, highlighting the differences among TICoder and the baseline methods, SimpleRAG and AllianceCoder.

SimpleRAG directly retrieves code snippets similar to the developer's requirement. However, as shown in Fig. \ref{fig_case}(c), it incorrectly uses the function ``\emph{\_hadoop\_streaming\_jar}'', while the correct callee should be ``\emph{get\_hadoop\_streaming\_jar}''. This indicates that retrieving code snippets based solely on similarity to the original requirement may fail to identify accurate and useful callee functions. 

On the other hand, AllianceCoder generates implementation details from the requirement and retrieves relevant APIs from the code repository. Yet, due to the lack of illustrative usage examples, it fails to correctly determine conditions such as ``\emph{no Hadoop streaming jar}'' as shown in Fig. \ref{fig_case}(d). 

In contrast, TICoder decomposes the developer's requirement, retrieves relevant callee functions, and provides corresponding usage patterns. This enables precise identification of the target callees and offers example-based guidance, allowing the LLM to effectively learn how to use them. As a result, TICoder achieves superior performance in repository-level code generation tasks.

\subsection{Threats to Validity}\label{threats}

\textbf{Threats to Internal Validity.}
Threats to internal validity mainly stem from the use of test cases, prompt design, and parameter settings in TICoder.
First, we incorporate test cases in both the planning stage and the enhanced code generation stage, inspired by test-driven development \citep{TDDBenchLLMCG, TDDCICG, TDDLLMCG}. However, in real-world repository-level code generation scenarios, high-quality test cases may not always be available, which could introduce threats to internal validity. Our approach highlights the benefit of incorporating test cases, in addition to NL requirements, for improving code generation. Meanwhile, recent advances in high-quality test case generation \citep{TDDLLMCG} may help mitigate this limitation.

Second, we adopt an iterative judge-and-reflection mechanism to refine planning steps. Although this design improves planning quality, its effectiveness may depend on the prompt templates. While our prompts are designed based on \emph{SRS} principles and prior work \citep{ClarifyGPT, SodaCoder}, they may not cover all aspects required for comprehensive planning and evaluation. To mitigate this risk, we sample 30 instances and involve two experienced programmers (each with over five years of professional experience) to iteratively refine the prompt templates for planning and judging, with the goal of improving their alignment with human programming practices. Additionally, not all tasks correspond to complex functions. \citet{DevEval} reported that 27\% of functions in DevEval are standalone and may not require planning or callee function retrieval. To mitigate this threat, we adopt an adaptive strategy in the planning stage, allowing the \emph{LLM-as-a-planner} to dynamically determine the number of implementation steps, where a single step implies no planning is needed.

Third, for dual-view similarity calculation, we conduct a limited grid search over the weight parameters on CoderEval (Python and Java), and select the best-performing configuration based on empirical observations. A more extensive exploration of parameter settings may further improve the performance of TICoder.

\textbf{Threats to External Validity.}
External validity concerns the generalizability of our findings. 
Our experiments are conducted on two widely-used benchmark datasets, CoderEval and DevEval, which may not fully represent all real-world repository-level code generation scenarios. In particular, these datasets may differ from industrial repositories in terms of scale, code quality, and dependency complexity. Additionally, our evaluation covers two programming languages, Python and Java. Future work could extend the evaluation to more diverse programming languages. 

Furthermore, although TICoder is designed to be model-agnostic, we evaluate it on three mainstream LLMs due to budget constraints. Given the rapid progress in LLMs, some LLMs not included in our study may achieve better performance. We will continue to track advancements in this area and incorporate more LLMs in future evaluations.

\textbf{Threats to Construct Validity.}
Construct validity concerns whether the evaluation metrics adequately reflect the desired properties of code generation systems. 
We adopt Pass@k as the primary evaluation metric, which measures functional correctness and is widely used in prior work \citep{AllianceCoder, RepoScope, DevEval, CoderEval}. However, Pass@k does not capture other important quality aspects, such as code readability, maintainability, or adherence to repository conventions. As a result, our evaluation may not fully reflect the overall quality of the generated code. Future work could incorporate human evaluation to provide a more comprehensive evaluation.

%
%

\section{Conclusion and Future Work}

In this paper, we present TICoder, a novel repository-level code generation framework that improves both planning and repository reuse through test-driven iterative planning and implementation-aware code reuse. Specifically, TICoder employs an iterative judge-and-reflection process guided by both requirements and test cases to refine implementation steps. Furthermore, TICoder introduces an implementation-aware reuse mechanism that integrates dual-view callee function retrieval and dual-stage usage pattern selection. In the retrieval stage, TICoder jointly considers functional and implementation similarity to identify relevant callee functions, while in the selection stage, it extracts representative usage patterns via structure-based clustering and perplexity-based filtering. Extensive experiments on two widely used benchmarks across multiple mainstream LLMs demonstrate that TICoder consistently outperforms SOTA baselines.

In addition to the threats discussed in Section~\ref{threats}, we plan to explore the following directions in future work: (1) investigating high-quality test case generation further to improve both planning steps and final code generation performance, (2) designing more refined prompt templates for planning, judging, reflection, and code generation, and (3) exploring deeper integration with agent-based frameworks, for example, exposing our iterative planning, dual-view retrieval, and dual-stage selection as modular tools for agents. 

\section{Data Availability Statement}
Our replication package is available at \url{https://doi.org/10.5281/zenodo.19342245} to facilitate future research.

\bibliographystyle{ACM-Reference-Format}
\bibliography{sample-base}


\end{document}